\documentclass[useAMS,usenatbib]{mn2e}
\usepackage{url,times,graphicx,amsmath,amsfonts,amssymb,aas_macros,color,epsfig}
\newcommand{\Tab}[1]{Table~\ref{#1}}

\newcommand{\Eq}[1]{Eq.~(\ref{#1})}
\newcommand{\Fig}[1]{Fig.~\ref{#1}}
\newcommand{\beq}{\begin{equation}}
\newcommand{\eeq}{\end{equation}}
\newcommand{\hMpc}{{\ifmmode{h^{-1}{\rm Mpc}}\else{$h^{-1}$Mpc}\fi}}
\newcommand{\hGpc}{{\ifmmode{h^{-1}{\rm Gpc}}\else{$h^{-1}$Gpc}\fi}}
\newcommand{\hkpc}{{\ifmmode{h^{-1}{\rm kpc}}\else{$h^{-1}$kpc}\fi}}
\newcommand{\hMsun}{{\ifmmode{h^{-1}{\rm {M_{\odot}}}}\else{$h^{-1}{\rm{M_{\odot}}}$}\fi}}
\def\beqa{\begin{eqnarray}}
\def\eeqa{\end{eqnarray}}
\def\hMpc{$h^{-1}\,{\rm Mpc}$}
\def\hkpc{$h^{-1}\,{\rm kpc}$}
\def\LCDM{\ensuremath{\Lambda}CDM}

\title[Vector dark energy and high-z massive clusters]
      {Vector dark energy and high-z massive clusters}
\author[Carlesi et al.]
       {Edoardo Carlesi,$^{1}$\thanks{E-mail: edoardo.carlesi@uam.es} 
	Alexander Knebe,$^{1}$ 
	Gustavo Yepes,$^{1}$ Stefan Gottl\"ober,$^{3}$ \newauthor
	Jose Beltr\'an Jim\'enez,$^{4,5}$ Antonio L. Maroto,$^{2}$ %, \newauthor
\\
\\
$^{1}$Departamento de F\'isica Te\'orica, Universidad Aut\'onoma de Madrid, 28049, Cantoblanco, Madrid, Spain\\
$^{2}$Departamento de F\'isica Te\'orica, Universidad Complutense de Madrid, 28040, Madrid, Spain \\
$^{3}$Leibniz Institut f\"ur Astrophysik, An der Sternwarte 16, 14482, Potsdam, Germany \\
$^{4}$Institute de Physique Th\'eorique, Universit\'e de Gen\`eve, 24 quai E. Ansermet, 1211 Gen\`eve, Switzerland\\
$^{5}$Institute of Theoretical Astrophysics, University of Oslo, 0315 Oslo, Norway
}

\begin{document}

\date{Accepted XXXX . Received XXXX; in original form XXXX}

\pagerange{\pageref{firstpage}--\pageref{lastpage}} \pubyear{2011}

\maketitle

\label{firstpage}

%%%%%%%%%%%%%%%%%%%%%%%%%%%%%%%%%%%%%%%%%%%%%%%%%%%
\begin{abstract}
  The detection of extremely massive clusters at $z>1$ such as SPT-CL J0546-5345,
  SPT-CL J2106-5844, and XMMU J2235.3-2557 has been considered by some authors
  as a challenge to the standard \LCDM$\;$cosmology.  In fact,
  assuming Gaussian initial conditions, the theoretical expectation of
  detecting such objects is as low as $\leq 1\%$.  In this \textit{Letter} we
  discuss the probability of the existence of such objects in the light of the Vector Dark
  Energy (VDE) paradigm, showing by means of a series of $N$-body
  simulations that chances of detection are substantially enhanced in
  this non-standard framework.
\end{abstract}
%%%%%%%%%%%%%%%%%%%%%%%%%%%%%%%%%%%%%%%%%%%%%%%%%%%


%%%%%%%%%%%%%%%%%%%%%%%%%%%%%%%%%%%%%%%%%%%%%%%%%%%
\section{Introduction} \label{sec:introduction}
%%%%%%%%%%%%%%%%%%%%%%%%%%%%%%%%%%%%%%%%%%%%%%%%%%%
Present day cosmology is still failing to explain satisfactorily the
nature of dark energy, which is supposed to dominate the energetic
content of the universe today and to be
responsible for the current accelerated expansion.  
In the standard \LCDM$\;$model, this cosmic acceleration
is generated by the presence of a cosmological constant. However, the
required value for that constant turns out to be tiny when compared to
the natural scale of gravity, namely the Planck scale. Thus, the
gravitational interaction would hence be described by two dimensional
constants differing by many orders of magnitude, and this poses a
problem of naturalness. This is the so-called ``cosmological constant
problem'' and it motivated to consider alternative explanations for
the current acceleration of the universe by either modifying the
gravitational interaction at large distances or introducing a new
dynamical field.
\\
Indeed, one of the main challenges of observational cosmology is
exactly to devise new tests which could help discriminating between
the constant or dynamic nature of dark energy.  In this regard,
several authors have recently pointed out that the observation of
extremely massive clusters at high redshift, such as SPT-CL J2106-5844
(\cite{Foley:2011}, $z\simeq1.18$, $M_{200} = (1.27 \pm 0.21) \times
10^{15} M_{\odot}$), SPT-CL J0546-5346 (\cite{Brodwin:2010}, $z\simeq1.07$, 
$M_{200} = (7.95\pm0.92)\times 10^{14} M_{\odot}$), 
and XMMU J2235.3-2557 (\cite{Jee:2009},
$z\simeq1.4$, $M_{200}=(7.3 \pm 1.3) \times 10^{14} M_{\odot}$) may
represent a major shortcoming of the \LCDM$\;$paradigm, where the
presence of such objects should be in principle strongly disfavoured
\citep[see, for example,][]{Baldi:2010,Mortonson:2011}.
\\
While, on the one hand, this tension could be solved keeping the
standard scenario and relaxing the assumption of Gaussianity in the
initial conditions (as proposed in \cite{Hoyle:2010} and
\cite{Enqvist:2011}), it could be as well possible to use this
observations as a constraint for different cosmological models.  In
this work we look at the VDE model, where the role of the dark energy
is played by a cosmic vector field
\citep{BeltranMaroto:2008}.  By means of a series of $N$-body
simulations, we study the large scale clustering properties of this
cosmology, computing the cumulative halo mass functions at different
redshifts and comparing them to the predictions of the standard model.
In this way, we are able to show that the VDE cosmology does indeed
predict a higher abundance of massive haloes at all redshifts, thus
enhancing the probability of observing such objects with respect to
$\Lambda$CDM.
%%%%%%%%%%%%%%%%%%%%%%%%%%%%%%%%%%%%%%%%%%%%%%%%%%%

%%%%%%%%%%%%%%%%%%%%%%%%%%%%%%%%%%%%%%%%%%%%%%%%%%% 
\section{Vector Dark Energy} \label{sec:model}
%%%%%%%%%%%%%%%%%%%%%%%%%%%%%%%%%%%%%%%%%%%%%%%%%%%

The action of the vector dark energy model (see \cite{BeltranMaroto:2008}) can be written as:
\begin{eqnarray}
S=\int d^4x \sqrt{-g}\left[-\frac{R}{16\pi G}
-\frac{1}{4}F_{\mu\nu}F^{\mu\nu}\right. \nonumber\\
\left.-\frac{1}{2}\left(\nabla_\mu
A^\mu\right)^2+ R_{\mu\nu}A^\mu A^\nu\right]. \label{CVaction}
\end{eqnarray}
where $R_{\mu\nu}$ is the Ricci tensor, $R=g^{\mu\nu}R_{\mu\nu}$ the 
scalar curvature and $F_{\mu\nu}=\partial_\mu A_\nu-\partial_\nu A_\mu$. 
This action can be interpreted as the Maxwell term for a vector field 
supplemented with a gauge-fixing term and an effective mass provided 
by the Ricci tensor. 
It is interesting to note that the vector sector 
has no free parameters nor potential terms, 
being $G$ the only dimensional constant of the theory.

For a homogeneous and isotropic universe described by the flat 
Friedmann-Lema\^itre-Robertson-Walker metric:
\begin{equation}
ds^2=dt^2-a(t)^2d\vec{x}^2
\end{equation}
we have
$A_\mu=(A_0(t),0,0,0)$ so that the corresponding equations read:
\begin{equation}
\ddot{A}_0+3H\dot{A}_0-3\left[2H^2+\dot{H}\right]A_0=0 \label{fieldeq0}
\end{equation}
\begin{equation}
H^2=\frac{8\pi G}{3} \left[\rho_R + \rho_M + \rho_A\right]
\end{equation}
with $H=\dot a/a$ the Hubble parameter and:
\begin{equation}
\rho_{A}=\frac{3}{2}H^2A_0^2+3HA_0\dot A_0-\frac{1}{2}\dot A_0^2
\end{equation}
the energy density associated to the vector field, while $rho_M$ and $rho_R$ are 
the matter and radiation densities.
During the radiation and matter eras in which the dark energy contribution was negligible, 
we can solve Eq. (\ref{fieldeq0}) with  $H=p/t$,
where $p=1/2$ for radiation and  $p=2/3$ for matter eras
respectively, that is equivalent to assume that $a\propto t^p$.
In that case, the general solution is:
\begin{equation}
A_0(t)=A_0^+t^{\alpha_+}+A_0^-t^{\alpha_-},\label{fieldsol}
\end{equation}
with $A_{0}^\pm$ constants of integration and $\alpha_{\pm}=-(1\pm
1)/4$ in the radiation era, and $\alpha_{\pm}=(-3\pm\sqrt{33})/6$
in the matter era. 
After dark energy starts dominating, the equation of state
abruptly falls towards $w_{DE}\rightarrow -\infty$ as the Universe
approaches $t_{end}$, and the equation
of state can cross the so-called phantom divide line (\cite{Nesseris:2006er}), 
so that we
can have $w_{DE}(z=0)<-1$.

Using the
growing mode solution in (\ref{fieldsol}) we can obtain the evolution for 
the energy density as:
\begin{equation}
\rho_{A}= \rho_{A0} (1+z)^\kappa,
\end{equation}
with $\kappa=4$ in the radiation era and $\kappa=(9-\sqrt{33})/2
\simeq -1.63$ in the matter era. Thus, the energy density of the
vector field scales like radiation at early times so that the ratio
$\rho_A/\rho_R$ is constant during such a period.  Moreover, the value
of the vector field $A_0$ during that era is also constant hence
making the cosmological evolution insensitive to the time at which we
impose the initial conditions (as long as they are set well inside the
radiation dominated epoch).  Also, such constant values are
$\rho_A/\rho_R\vert_{\rm early}\simeq 10^{-6}$ and $A_0^{\rm
  early}\simeq 10^{-4} M_p$ which are values that can arise naturally
during the early universe, for instance, as quantum
fluctuations. Furthermore, they do not need the introduction of any
unnatural scale, thus, alleviating the naturalness or coincidence
problem.  On the other hand, when the Universe enters the era of
matter domination, $\rho_A$ starts growing relative to $\rho_M$
eventually overcoming it at some point so that the dark energy vector
field becomes the dominant component.

\begin{figure}
\vspace{0.2cm}
\begin{center} 
\includegraphics[width=6cm]{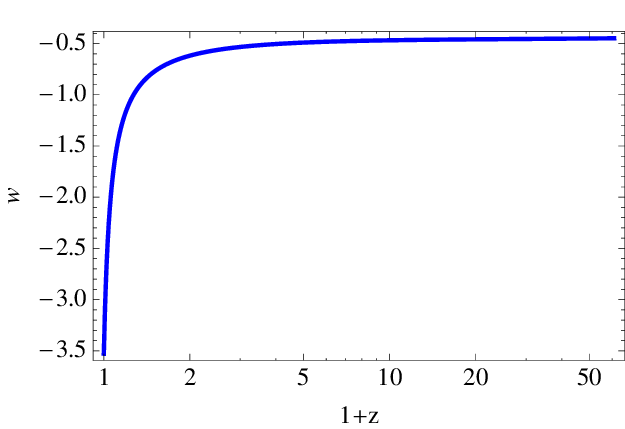}
\caption{Equation of state of the vector dark energy model for the best fit to SNIa data.}
\label{EoS}
\end{center}
\end{figure}

Once the present value of the Hubble parameter $H_0$ and the constant
$A_0^{early}$ during radiation (which fixes the total amount of matter
$\Omega_M$) are specified, the model is completely determined. In
other words, this model contains the same number of parameters as
\LCDM, i.e. the minimum number of parameters of any cosmological model
with dark energy. Notice however, that in the VDE model the present
value of the equation of state parameter $w_0=-3.53$ is radically
different from that of a cosmological constant (cf. \Fig{EoS}, where the redshift
evolution of $\omega(z)$ is shown our the range of our simulations).
Despite this fact, VDE is able to fit supernovae and CMB data with
comparable goodness to \LCDM\ \citep{BeltranMaroto:2008, BLM:2009}.
%%%%%%%%%%%%%%%%%%%%%%%%%%%%%%%%%%%%%%%%%%%%%%%%%%%

%%%%%%%%%%%%%%%%%%%%%%%%%%%%%%%%%%%%%%%%%%%%%%%%%%%
\section{The Data} \label{sec:setup}
%%%%%%%%%%%%%%%%%%%%%%%%%%%%%%%%%%%%%%%%%%%%%%%%%%%
\subsection{Simulations}
We wanted to estimate the probability of finding massive clusters at
$z>1$ in the VDE scenario compared to the \LCDM$\;$one by means of CDM
only $N$-body simulations.  For this purpose, we chose to use a
suitably modified version of the publicly available GADGET-2 tree-PM
code \citep{Springel:2005}, which had to take into account the
different expansion history that characterizes the two cosmologies.
In \Tab{tab:cosmoparam} we show the cosmological parameters used in
the different simulations. 
For the VDE model, we have used the value of $\Omega_M$ 
provided by the best fit SNIa and then we have fitted WMAP7
in order to obtain the remaining cosmological parameters.
$w_0$ denotes the present value of the equation 
of state parameter of dark energy.
For \LCDM\ we used the
Multidark Simulation \citep{Prada:2011} cosmological parameters
with a WMAP7 $\sigma_8$ normalization \citep{Wmap:2011}.
We also introduced a so called \LCDM-vde model, which is a standard \LCDM\ one implementing 
the same $\Omega_M$ and $\sigma_8$ as VDE. Although this cosmology is non-viable and ruled out by
experimental data, its study allows us to disentangle
and higlight the effect of the increased matter density and of matter perturbations normalization 
on our findings.
\\
We chose to run a total of eight $512^3$ particles simulations
summarized in \Tab{tab:settings} and explained below:

\begin{itemize}
\item a VDE (and a \LCDM\ started with the same seed for
  the phases of the initial conditions) simulation in a
500 \hMpc$\;$box,
\item a second VDE (and again corresponding \LCDM)  simulation in a 1 \hGpc$\;$box,
\item two more VDE simulations with a different random seed, one in a 500 \hMpc$\;$
and one in a 1 \hGpc$\;$box, as a check for the influence of cosmic variance,
\item two \LCDM-vde simulations in a 500 \hMpc\ and a 1000 \hMpc\ box.
\end{itemize}

The full set of simulations will be presented and analyzed in an
upcoming companion paper; in this work, instead, we chose to focus on
some of them only in order to gather information on large scale
clustering in the two cosmologies.  The use of the same initial seed
in the coupled \LCDM-VDE simulations allows us to directly compare the
structures identified by the halo finder, which are supposed to form
at the same points corresponding to the overdensity peaks formed from
the initial Gaussian density field.

\begin{table}
\caption{Cosmological parameters for \LCDM\, \LCDM-vde and VDE.
 }

\begin{center}
\begin{tabular}{cccccc}
\hline
Model & $\Omega_{m}$ & $\Omega_{de}$ &$w_0$& $\sigma_8$ & h  \\
\hline
\LCDM & 0.27 & 0.73 &-1& 0.8 & 0.7 \\
\LCDM-vde & 0.388 & 0.612 &-1& 0.83 & 0.7 \\
VDE & 0.388 & 0.612 & -3.53& 0.83 & 0.62 \\
\hline
\end{tabular}
\end{center}
\label{tab:cosmoparam}
\end{table}

\begin{table}
\caption{$N$-body settings used for the GADGET-2 simulations, 
  the two 500\hMpc$\;$and the two 1\hGpc$\;$have the same initial random seed
  and starting redshift $z_{\rm start}=60$
  in order to allow for a direct comparison of the halo
  properties. The number of particles in each was fixed at $512^3$.The box size $B$ is given in \hMsun\ and the particle
  mass in \hMsun. The cosmology refers back to the parameters listed
  in \Tab{tab:cosmoparam}.}
\begin{center}
\begin{tabular}{lcc}
\hline
Simulation &  $B$ & $m_{p}$  \\
\hline
$2\times$VDE-0.5           &  500 & $1.00\times 10^{11}$ \\
$2\times$VDE-1           &  1000 & $8.02\times 10^{11}$ \\
$\Lambda$CDM-0.5   &  500   & $6.95\times 10^{10}$ \\
$\Lambda$CDM-1   &  1000 &  $5.55\times 10^{11}$ \\
$\Lambda$CDM-0.5vde   &  500 & $1.00\times 10^{11}$ \\
$\Lambda$CDM-1vde   &  1000 & $8.02\times 10^{11}$ \\
\hline
\end{tabular}
\end{center}
\label{tab:settings}
\end{table}
As a final remark, we underline here that the choice of the boxes was
made in order to allow the study of clustering on larger scales,
without particular emphasis on the low mass results, e.g. objects with
$M<10^{14}$\hMsun.  This means that even though our halo finder
has been able to identify objects down to $\sim
10^{12} \hMsun$ in the 500\hMpc$\;$box and $\sim 10^{13}$
\hMsun$\;$in the 1\hGpc$\;$one (which correspond to a
lower limit of 20 particles), we are not comparing the
mass spectrum at this far end.  Therefore, since we are only
interested in studying the behaviour of the mass function of these
models at the very high mass end, in the following section we will
refer mostly to the \LCDM-1, \LCDM-vde and VDE-1 simulations, where we have a
larger statistics for the supercluster scales.

\subsection{Halo Finding}\label{sec:analysis}
In order to identify halos in our simulation we have run the
MPI+OpenMP hybrid halo finder \texttt{AHF}\footnote{\texttt{AMIGA} halo
finder, to be downloaded freely from
\texttt{http://www.popia.ft.uam.es/AMIGA}} described in detail in
\cite{Knollmann:2009}. \texttt{AHF} is an improvement of the
\texttt{MHF} halo finder \citep{Gill:2004}, which locates local
overdensities in an adaptively smoothed density field as prospective
halo centres. The local potential minima are computed for each of
these density peaks and the gravitationally bound particles are
determined. Only peaks with at least 20 bound particles are considered
as haloes and retained for further analysis, even though here we focus
on the most massive objects only.

The mass of each halo is then computed via the equation

\begin{equation}
M(r) = \frac{4 \pi}{3} \Delta \rho_{c} r^{3}
\end{equation}
where we applied $\Delta=200$ as the overdensity threshold. 
Using this relation, particular care has to be taken when considering the definition of the critical density
 
\begin{equation}
\rho_{c} = \frac{3 H^2}{8 \pi G} 
\end{equation}
because it involves the Hubble parameter, that differs substantially
at all redshifts in the two models.  This means that, identifying the
halo masses, we have to take into account the fact that the value of
$\rho_c$ changes from \LCDM$\;$and VDE.  This has been incorporated
into and taken care of in the latest version of \texttt{AHF} where
$H_{VDE}(z)$ is being read in from a precomputed table.

We would like to mention that we checked that the objects obtained by this (virial) definition are in fact in equilibrium. To this extent we studied the ratio between two times kinetic over potential energy $2T/|U|$ confirming that at each redshift under investigation here this relation is equally well fulfilled for the \LCDM\ and -- more importantly -- the VDE simulations. We therefore conclude that our adopted method to define halo mass in the VDE model leads to unbiased results.

%%%%%%%%%%%%%%%%%%%%%%%%%%%%%%%%%%%%%%%%%%%%%%%%%%%

%%%%%%%%%%%%%%%%%%%%%%%%%%%%%%%%%%%%%%%%%%%%%%%%%%%
\section{The Results} \label{sec:results}
%%%%%%%%%%%%%%%%%%%%%%%%%%%%%%%%%%%%%%%%%%%%%%%%%%%
\begin{figure*}\begin{center} 
\includegraphics[angle=270,width=13cm]{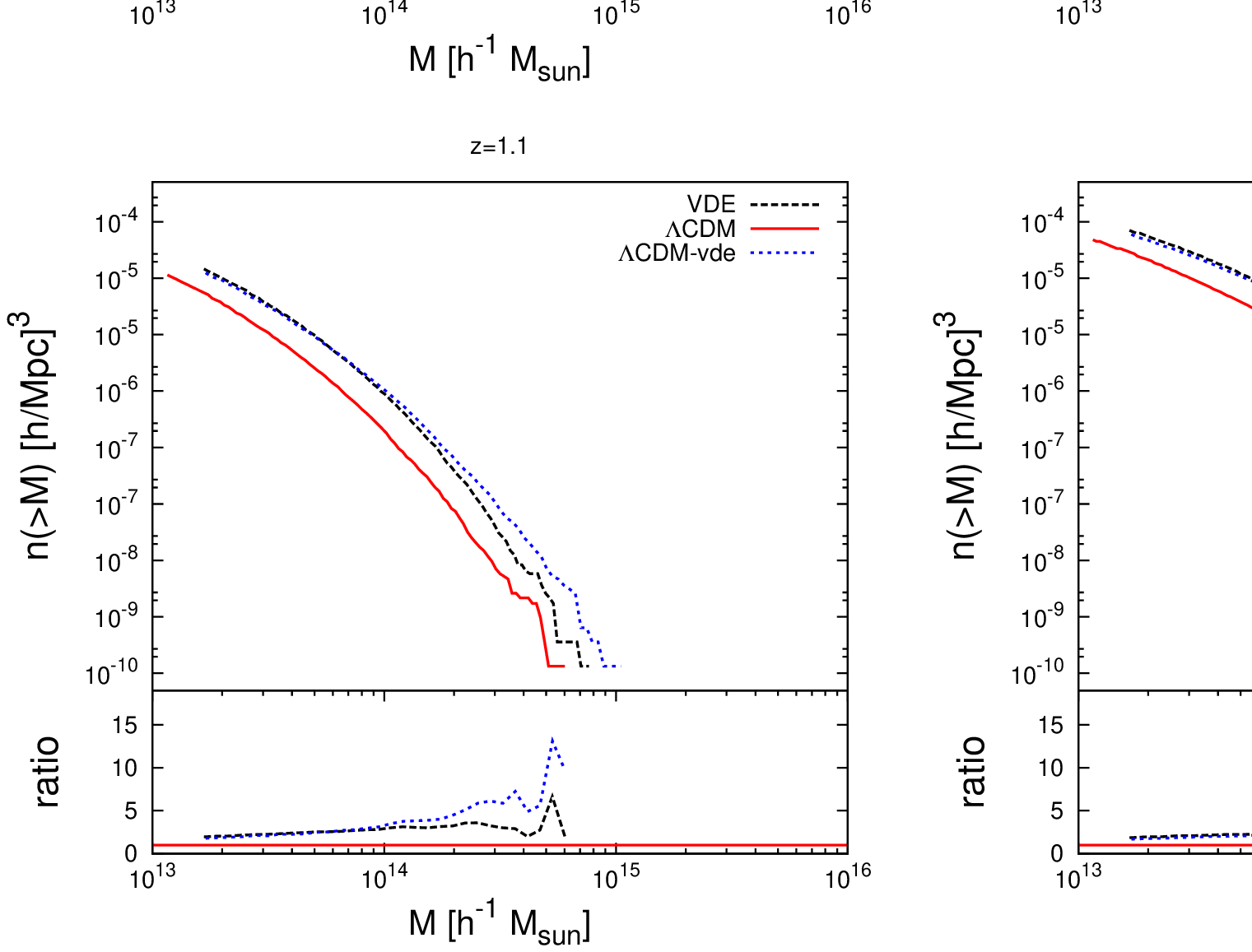}
\caption{Mass functions (and their ratios) 
as computed for the VDE-1, \LCDM-1 and \LCDM-1vde simulations at $z=1.4, 1.2, 1.1,$ and $0$. These redshifts have been chosen in order to overlap with the aforementioned observed massive clusters.
}
\label{img:massfunctions}
\end{center}\end{figure*}

\subsection{Mass Function}
With the halo catalogues at our disposal, we computed the cumulative
mass functions $n(>M)$ at various redshifts. We show in
\Fig{img:massfunctions} the results for the 1\hGpc\ simulations at redshifts
$z=1.4$, $z=1.2$, $z=1.1$ and $z=0$. This plot is accompanied by
\Tab{tab:massivehaloes} where we list the masses of the most
massive haloes found in each model and the redshifts under consideration.

We notice that the mass function for $M > 10^{14} \hMsun$ is
several times larger in VDE than in \LCDM$\;$at all redshifts,
thus increasing significantly the number of higher mass haloes in this non-standard cosmological model. 
In particular, at the high-mass end the VDE mass function is about three
times larger at the relevant redshifts $z=1.4, 1.2,$ and $1.1$ -- and even larger at today's time.

In order to verify that this feature of the VDE model is not a simple reflection of cosmic variance (which should affect in particular the high mass end, where the statistics is smaller) we compared the results presented in \Fig{img:massfunctions} to the mass functions of the set of two additional simulations started from a different random seed for the initial conditions.  In fact, the VDE cumulative
mass function turns out to outnumber the \LCDM$\;$ one by the same
factor at all redshifts in these two test runs, too.

An interesting
remark we would like to add here, is that the physical mass 
(obtained dividing by the corresponding $h$ values the values quoted in \hMsun units)
of the largest haloes in the VDE-1 simulation at $z=1.4$, $z=1.2$ and $z=1.1$ are perfectly
compatible with the ones of the above cited clusters, whereas the
corresponding \LCDM$\;$ candidates are outside the $2\sigma$ compatibility
level. And again, similar massive clusters have also
been found in the duplicate VDE-1 simulation with a different initial
seed.

\begin{table}
\caption{The most massive halo found in the three 1\hGpc$\;$
  simulations (in units of $10^{14}$\hMsun) 
  as a function of redshift.
}
\begin{center}
\begin{tabular}{cccc}
\hline
z & \LCDM-1 & VDE-1 & \LCDM-1vde \\
\hline
1.4 & 4.16 & 5.63  & 6.47 \\
1.2 & 5.13 & 6.51  & 8.16 \\
1.1 & 6.01 & 7.63  & 10.2 \\
0   & 18.1 & 31.6  & 35.1 \\
\hline
\end{tabular}
\end{center}
\label{tab:massivehaloes}
\end{table}

As an additional note, we observe that the \LCDM-1vde simulations yields a mass
function is almost indistinguishable from the VDE one for $M<10^{14}\times\hMsun$ whereas it 
is a factor of $\sim3$ higher than the VDE one in the high mass range.
This is a clear indication that the higher normalization of the matter
fluctuations and, most important, the higher value of $\Omega_M$ act as the main sources of the enhancement of
clustering found in VDE-1 and VDE-0.5.  On the one hand, this complicates the issue of model
selection, since (although disfavoured by the WMAP7 data) we could
invoke a bigger $\Omega_M$ or a higher $\sigma_8$ normalization at $z=0$ for \LCDM\ to
explain the current tension with the high-$z$ massive clusters
observations. On the other hand, the distinct expansion history that
characterizes and differentiates between the two \LCDM\ and VDE models would still
leave a clear imprint on structure formation at different times,
which could be detected e.g. measuring $\sigma_8$'s dependence on the
redshift.  Such a test would indeed provide invaluable information for
the study of \LCDM$\;$and for any cosmological model beyond it such as
VDE.

%%%%%%%%%%%%%%%%%%%%%%%%%%%%%%%%%%%%%%%%%%%%%%%%%%%
\subsection{Probability} \label{sec:numest}
%%%%%%%%%%%%%%%%%%%%%%%%%%%%%%%%%%%%%%%%%%%%%%%%%%%

\begin{figure}
\vspace{0.2cm}
\begin{center} 
\includegraphics[width=5cm, angle=270]{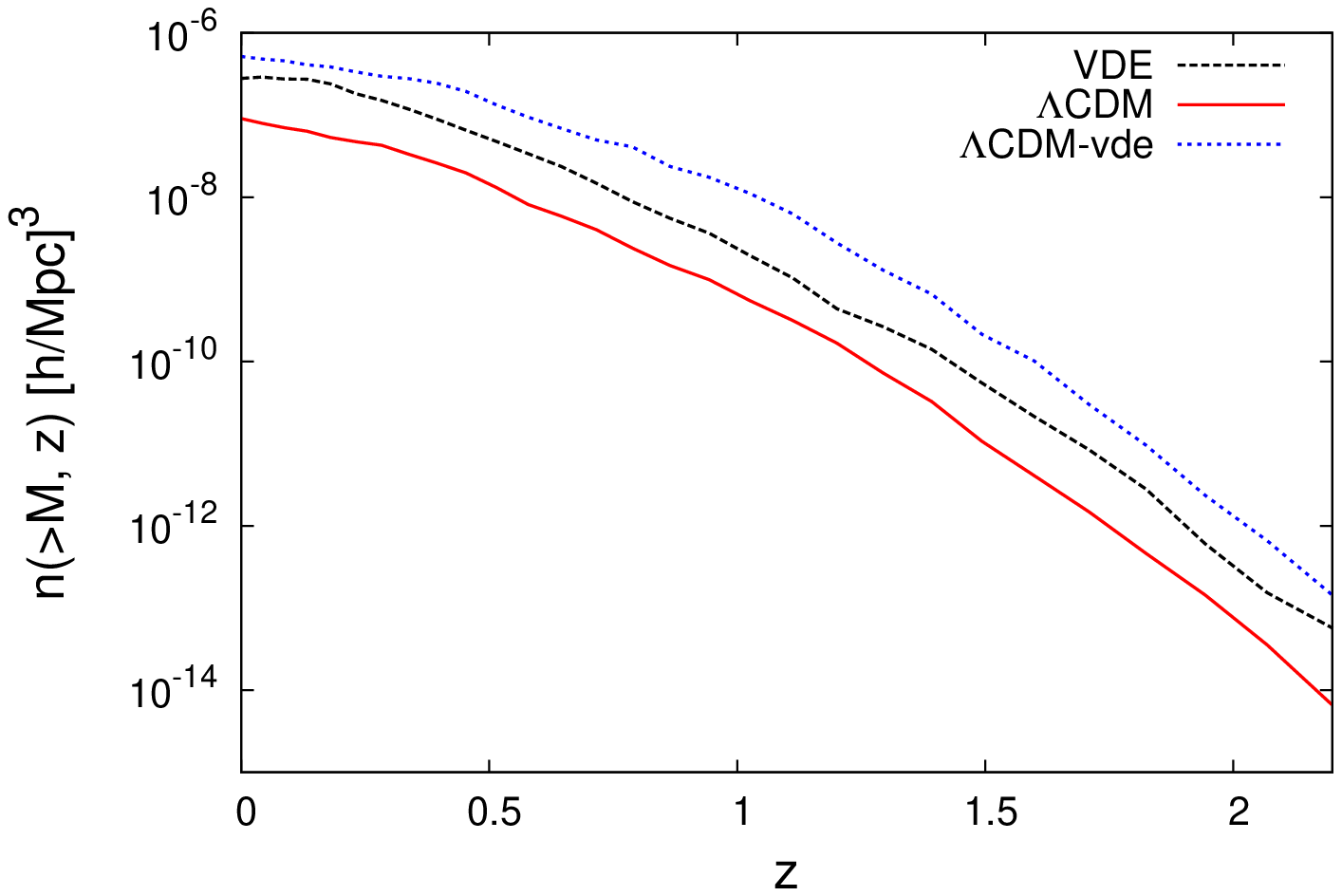}
\caption{Theoretical cumulative number densities 
of objects with $M>5\times 10^{14}M_{\odot}$ for VDE and \LCDM.}
\label{fig:nd}
\end{center}
\end{figure}

In order to provide a more quantitative estimate of the the relative probability
of observationally detecting such massive clusters  
we used $n(>M,z)$ -- the expected cumulative number density
of objects above a threshold mass $M$ as a function of redshift as given by our simulations -- and integrated it over the comoving volume $V_c$ of the survey

\begin{equation}\label{eqn:ninteg}
N(>M) = \int_{\Delta z, \Omega_{\rm survey}} n(>M, z) d V_c(z) 
\end{equation}
where $\Delta z$ and $\Omega_{\rm survey}$ are the redshift interval and the fraction of the sky
covered by the survey to which we want to compare our theoretical expectations.

While $n(>M,z)$ can be readily calculated in \LCDM$\;$cosmologies \citep[e.g.][]{PressSchechter:1974,ShethTormen:1999,Jenkins:2001,Tinker:2008}, 
in VDE we have to devise a strategy to compute it based upon our numerical results only by adjusting the formula of \citet{ShethTormen:1999}:

\begin{itemize}
\item we calculated the cumulative number densities in the desired redshift intervals $\Delta z$ based upon our simulation data,
\item we adjusted the parameters of the Sheth-Tormen mass function fitting the numerical cumulative number densities  
derived from the VDE-1 and VDE-0.5 simulations,
\item we used these best fit estimates to analytically compute $n(>M, z)$ now having access to masses outside our numerically limited range to be used with \Eq{eqn:ninteg}.
\end{itemize}

The results of the numerical integration over the comoving volumes
(obtained using the limits quoted in the observational papers by \cite{Jee:2009}, \cite{Brodwin:2010} and \cite{Foley:2011})
are listed in \Tab{tab:ndensity} for the VDE, \LCDM-vde and \LCDM\ model. We can clearly see that the chances are substantially larger to find such massive objects as the ones observed by \cite{Jee:2009}, \cite{Brodwin:2010} and \cite{Foley:2011} in VDE than in \LCDM. We complement these results with \Fig{fig:nd} where we plot the abundance evolution of clusters with mass $M>5\times 10^{14}$\hMsun\ computed with above described procedure. This plot confirms our previous analysis of the mass functions and shows that the 
expectation of massive objects is enhanced in VDE by a factor $\sim 3$ to $\sim 10$ at all redshifts, 
a factor which is even higher for \LCDM-vde.
We would like to remark here that while our \LCDM$\;$ estimate for the third 
cluster is in agreement with the result quoted by 
\cite{Jee:2009} (obtained using our same approach), the same calculation done for the first one
leads to an estimate substantially smaller than the one quoted by \cite{Foley:2011}, calculated using a 
Monte Carlo technique. However, this does not affect our conclusions, which are based on 
the comparison of results obtained in a consistent manner for the two models.

\begin{table}
\caption{Expected number of objects $N(>M)$ in excess of mass $M$ and inside a certain (comoving) volume in the \LCDM$\;$ and VDE for different mass thresholds and survey volumes.
Solid angles $\Omega$ are measured in deg$^2$ and masses are measured in $10^{14}$\hMsun.}
\begin{center}
\begin{tabular}{cccccc}
\hline
$M$ & $\Delta z$ & $\Omega_{\rm survey}$ & $N_{\Lambda\rm CDM}$ & $N_{\rm VDE}$ & $N_{\Lambda\rm-VDE}$ \\
\hline
$>10$ & $>1$   	& 2500 & 0.007  & 0.02  & 0.04 \\
$>7$  & $>1$   	& 2500 & 0.03   & 0.31  & 0.56  \\
$>5$  & $1.38-2.2$  & 11& 0.005  & 0.06  &  0.07 \\
\hline
\end{tabular}
\end{center}
\label{tab:ndensity}
\end{table}

%%%%%%%%%%%%%%%%%%%%%%%%%%%%%%%%%%%%%%%%%%%%%%%%%%%
\section{Conclusions} \label{sec:conclusions}
%%%%%%%%%%%%%%%%%%%%%%%%%%%%%%%%%%%%%%%%%%%%%%%%%%%
The observation of massive clusters at $z>1$ provides an additional,
useful test for \LCDM$\;$and other cosmological models beyond the standard
paradigm.  In this \textit{Letter} we have shown that the Vector Dark
Energy (VDE) scenario \citep{BeltranMaroto:2008} might account for
such observations better than the \LCDM$\;$concordance model, since
the relative abundance of extremely massive cluster is at all
redshifts higher in this non-standard cosmology. 
Computing the cumulative number density 
at different redshifts, we estimated that the expected number of
massive clusters is enhanced in VDE by at least a factor of $\sim 3$ for the $M=10^{15}\times\hMsun$ cluster and 
a factor of $\sim10$ in the other two cases.
Of course, these
results might as well simply point in the direction of modifying the
standard paradigm, for example including non-Gaussianities in the
initial conditions or either using a higher $\sigma_8$ or $\Omega_M$ value for the
\LCDM\, as the comparison to the \LCDM-vde model seems to suggest.

Nonetheless, this first results on the large
scale clustering in the case of VDE cosmology point in the right
direction, significantly enhancing the probability of producing
extremely massive clusters at high redshift as recent observations
seem to require.
%%%%%%%%%%%%%%%%%%%%%%%%%%%%%%%%%%%%%%%%%%%%%%%%%%%

%%%%%%%%%%%%%%%%%%%%%%%%%%%%%%%%%%%%%%%%%%%%%%%%%%%
\section*{Acknowledgements}
EC is supported by the MareNostrum project funded by the Spanish
Ministerio de Ciencia e Innovacion (MICINN) under grant
no. AYA2009-13875-C03-02 and MultiDark Consolider project under grant
CSD2009-00064. EC also acknowledges partial support from the European 
Union FP7 ITN INVISIBLES (Marie Curie Actions, PITN- GA-2011- 289442).
AK acknowledges support by the MICINN's Ramon y Cajal
programme as well as the grants AYA 2009-13875-C03-02,
AYA2009-12792-C03-03, CSD2009-00064, and CAM S2009/ESP-1496. 
JBJ is supported by the
Ministerio de Educaci\'on under the postdoctoral contract EX2009-0305
and also wishes to acknowledge support from the Norwegian Research
Council under the YGGDRASIL programme 2009-2010 and the NILS mobility
project grant UCM-EEA-ABEL-03-2010.  We also acknowledge support from
MICINN (Spain) project numbers FIS 2008-01323, FPA 2008-00592 and 
CAM/UCM 910309.

%%%%%%%%%%%%%%%%%%%%%%%%%%%%%%%%%%%%%%%%%%%%%%%%%% 

\bibliographystyle{mn2e}
\bibliography{biblio}

\begin{thebibliography}{}

\bibitem[\protect\citeauthoryear{{Baldi} \& {Pettorino}}{{Baldi} \&
  {Pettorino}}{2011}]{Baldi:2010}
{Baldi} M.,  {Pettorino} V.,  2011, \mnras, 412, L1

\bibitem[\protect\citeauthoryear{{Beltr{\'a}n Jim{\'e}nez}, {Lazkoz} \&
  {Maroto}}{{Beltr{\'a}n Jim{\'e}nez} et~al.}{2009}]{BLM:2009}
{Beltr{\'a}n Jim{\'e}nez} J.,  {Lazkoz} R.,    {Maroto} A.~L.,  2009, \prd, 80,
  023004

\bibitem[\protect\citeauthoryear{{Beltr{\'a}n Jim{\'e}nez} \&
  {Maroto}}{{Beltr{\'a}n Jim{\'e}nez} \& {Maroto}}{2008}]{BeltranMaroto:2008}
{Beltr{\'a}n Jim{\'e}nez} J.,  {Maroto} A.~L.,  2008, \prd, 78, 063005

\bibitem[\protect\citeauthoryear{{Brodwin} et~al.,}{{Brodwin}
  et~al.}{2010}]{Brodwin:2010}
{Brodwin} M.,  et~al., 2010, \apj, 721, 90

\bibitem[\protect\citeauthoryear{{Enqvist}, {Hotchkiss} \& {Taanila}}{{Enqvist}
  et~al.}{2011}]{Enqvist:2011}
{Enqvist} K.,  {Hotchkiss} S.,    {Taanila} O.,  2011, JCAP, 4

\bibitem[\protect\citeauthoryear{Foley et~al.,}{Foley
  et~al.}{2011}]{Foley:2011}
Foley R.,  et~al., 2011, \apj, 731, 86

\bibitem[\protect\citeauthoryear{{Gill}, {Knebe} \& {Gibson}}{{Gill}
  et~al.}{2004}]{Gill:2004}
{Gill} S.~P.~D.,  {Knebe} A.,    {Gibson} B.~K.,  2004, \mnras, 351, 399

\bibitem[\protect\citeauthoryear{{Hoyle}, {Jimenez} \& {Verde}}{{Hoyle}
  et~al.}{2011}]{Hoyle:2010}
{Hoyle} B.,  {Jimenez} R.,    {Verde} L.,  2011, \prd, 83, 103502

\bibitem[\protect\citeauthoryear{{Jee} et~al.,}{{Jee}  et~al.}{2009}]{Jee:2009}
{Jee} M.,  et~al., 2009, \apj, 704, 672

\bibitem[\protect\citeauthoryear{{Jenkins} et~al.,}{{Jenkins}
  et~al.}{2001}]{Jenkins:2001}
{Jenkins} A.,  et~al., 2001, \mnras, 321, 372

\bibitem[\protect\citeauthoryear{{Knollmann} \& {Knebe}}{{Knollmann} \&
  {Knebe}}{2009}]{Knollmann:2009}
{Knollmann} S.~R.,  {Knebe} A.,  2009, \apjs, 182, 608

\bibitem[\protect\citeauthoryear{{Larson} et~al.,}{{Larson}
  et~al.}{2011}]{Wmap:2011}
{Larson} D.,  et~al., 2011, \apjs, 192, 16

\bibitem[\protect\citeauthoryear{{Mortonson}, {Hu} \& {Huterer}}{{Mortonson}
  et~al.}{2011}]{Mortonson:2011}
{Mortonson} M.~J.,  {Hu} W.,    {Huterer} D.,  2011, \prd, 83, 023015

\bibitem[\protect\citeauthoryear{Nesseris \& Perivolaropoulos}{Nesseris \&
  Perivolaropoulos}{2007}]{Nesseris:2006er}
Nesseris S.,  Perivolaropoulos L.,  2007, JCAP, 0701, 018

\bibitem[\protect\citeauthoryear{{Prada}, {Klypin}, {Cuesta}, {Betancort-Rijo}
  \& {Primack}}{{Prada} et~al.}{2011}]{Prada:2011}
{Prada} F.,  {Klypin} A.~A.,  {Cuesta} A.~J.,  {Betancort-Rijo} J.~E.,
  {Primack} J.,  2011, ArXiv e-prints

\bibitem[\protect\citeauthoryear{{Press} \& {Schechter}}{{Press} \&
  {Schechter}}{1974}]{PressSchechter:1974}
{Press} W.~H.,  {Schechter} P.,  1974, \apj, 187, 425

\bibitem[\protect\citeauthoryear{{Sheth} \& {Tormen}}{{Sheth} \&
  {Tormen}}{1999}]{ShethTormen:1999}
{Sheth} R.~K.,  {Tormen} G.,  1999, \mnras, 308, 119

\bibitem[\protect\citeauthoryear{{Springel}}{{Springel}}{2005}]{Springel:2005}
{Springel} V.,  2005, \mnras, 364, 1105

\bibitem[\protect\citeauthoryear{{Tinker}, {Kravtsov}, {Klypin}, {Abazajian},
  {Warren}, {Yepes}, {Gottl{\"o}ber} \& {Holz}}{{Tinker}
  et~al.}{2008}]{Tinker:2008}
{Tinker} J.,  {Kravtsov} A.~V.,  {Klypin} A.,  {Abazajian} K.,  {Warren} M.,
  {Yepes} G.,  {Gottl{\"o}ber} S.,    {Holz} D.~E.,  2008, \apj, 688, 709

\end{thebibliography}

\bsp

\label{lastpage}

\end{document}